\documentclass[sigconf,review=False]{acmart}

\usepackage{array}
\usepackage{graphicx}
\usepackage{clrscode}
\usepackage{balance}
\usepackage{subfigure}
\usepackage{multirow}
\usepackage{booktabs}
\usepackage{adjustbox}
\usepackage[justification=justified,skip=5pt]{caption}
\usepackage{float}
\usepackage{color}
\usepackage{soul}
\usepackage{mdframed}
\usepackage{amsopn}
\usepackage{mathrsfs}
\usepackage{mathtools}
\usepackage{amsmath}
\usepackage{arydshln}
\usepackage{hyperref}
\usepackage{multicol}
\usepackage{blkarray}
\usepackage{enumerate}
\usepackage{courier}
\usepackage{rotating}
\usepackage{booktabs}
\usepackage{diagbox}
\usepackage{fancybox}
\usepackage{minibox}
\usepackage{cases}
\usepackage{bm}
\usepackage{enumitem}
\usepackage[lined, ruled, commentsnumbered, linesnumbered]{algorithm2e} %
\usepackage{enumitem}
\setlist[itemize]{leftmargin=*}

\DeclareMathOperator*{\argmax}{arg\,max}

\AtBeginDocument{%
  \providecommand\BibTeX{{%
    \normalfont B\kern-0.5em{\scshape i\kern-0.25em b}\kern-0.8em\TeX}}}

\copyrightyear{2022} 
\acmYear{2022} 
\setcopyright{acmcopyright}\acmConference[WSDM '22]{Proceedings of the Fifteenth ACM International Conference on Web Search and Data Mining}{February 21--25, 2022}{Tempe, AZ, USA}
\acmBooktitle{Proceedings of the Fifteenth ACM International Conference on Web Search and Data Mining (WSDM '22), February 21--25, 2022, Tempe, AZ, USA}
\acmPrice{15.00}
\acmDOI{10.1145/3488560.3498487}
\acmISBN{978-1-4503-9132-0/22/02}

\begin{document}
\fancyhead{}
\title{Toward Pareto Efficient Fairness-Utility Trade-off in Recommendation through Reinforcement Learning}

\author{Yingqiang Ge$^{\dagger}$, Xiaoting Zhao${^\ast}$, Lucia Yu${^\ast}$, Saurabh Paul${^\ast}$, }
\author{Diane Hu${^\ast}$, Chu-Cheng Hsieh${^\ast}$, Yongfeng Zhang${^\dagger}$}
\affiliation{%
  \institution{$^{\dagger}$Rutgers University \qquad ${^\ast}$ 
Etsy Inc. }
\country{}
}
\email{yingqiang.ge@rutgers.edu, {xzhao, lyu, spaul, dhu, chsieh}@etsy.com, 
yongfeng.zhang@rutgers.edu}

\renewcommand{\authors}{Yingqiang Ge, Xiaoting Zhao, Lucia Yu, Saurabh Paul, Diane Hu, Chu-Cheng Hsieh, Yongfeng Zhang}









\begin{abstract}
The issue of fairness in recommendation is becoming increasingly essential as Recommender Systems (RS) touch and influence more and more people in their daily lives.
In fairness-aware recommendation, most of the existing algorithmic approaches mainly aim at solving a constrained optimization problem by imposing a constraint on the level of fairness while optimizing the main recommendation objective, e.g., click through rate  (CTR).
While this alleviates the impact of unfair recommendations, the expected return of an approach may significantly compromise the recommendation accuracy due to the inherent trade-off between fairness and utility. 
This motivates us to deal with these conflicting objectives and explore the optimal trade-off between them in recommendation.
One conspicuous approach is to seek a \textit{Pareto efficient/optimal} solution to guarantee optimal compromises between utility and fairness.
Moreover, considering the needs of real-world e-commerce platforms, it would be more desirable if we can generalize the whole \textit{Pareto Frontier}, so that the decision-makers can specify any preference of one objective over another based on their current business needs.
Therefore, in this work, we propose a fairness-aware recommendation framework using \textit{multi-objective reinforcement learning} (MORL), called MoFIR (pronounced ``\textbf{more fair}''), which is able to learn a single parametric representation for optimal recommendation policies over the space of all possible preferences.
Specially, we modify traditional Deep Deterministic Policy Gradient (DDPG) by introducing \textit{conditioned network} (CN) into it, which conditions the networks directly on these preferences and outputs Q-value-vectors.
Experiments on several real-world recommendation datasets verify the superiority of our framework on both fairness metrics and recommendation measures when compared with all other baselines.
We also extract the approximate Pareto Frontier on real-world datasets generated by MoFIR and compare to state-of-the-art fairness methods.

\end{abstract}





\maketitle

\section{Introduction}
Personalized recommender systems (RS), which are extensively employed in e-commerce platforms, have been acknowledged for their capacity to deliver high-quality services that bridge the gap between products and customers \cite{chen2022graph,ge2020learning,tan2021counterfactual,xu2021causal}.
Despite these huge advantages, several recent studies also raised concerns that RS may be vulnerable to algorithmic bias in several aspects, which may result in detrimental consequences for underrepresented or disadvantaged groups~\cite{Geyik2019,Zhu2018,singh2018fairness,li2021tutorial}.
For example, the ``Matthew Effect'' becomes increasingly evident in RS, which creates a huge disparity in the exposure of the producers/products in real-world recommendation systems \cite{10.1145/3292500.3330707,10.1145/3437963.3441824,ge2020understanding}.
Fortunately, these concerns about algorithmic fairness have resulted in a resurgence of interest to develop fairness-aware recommendation models to ensure such models do not become a source of unfair discrimination in recommendation \cite{li2021cikm,li2021towards,ge2021towards,fu2020fairness}.

In the area of fairness-aware recommendation, the methods can be roughly divided into three categories: pre-processing, in-processing and post-processing algorithms \cite{10.1145/3404835.3462807,li2021tutorial}.
Pre-processing methods usually aim to remove bias in data, e.g., sampling from data to  cover items of all groups or balancing data to increase coverage of minority groups.
In-processing methods aim at encoding fairness as part of the objective function, while post-processing methods tend to modify the presentations of the results.
Even though all of them could successfully alleviate the impact of unfair recommendations to some extent, the expected return of an approach may significantly compromise the recommendation accuracy due to the inherent trade-off between fairness and utility, which has been demonstrated by several recent work both empirically and theoretically~\cite{kamani2021pareto,kearns2019ethical,NEURIPS2018_8e038477,10.5555/3294771.3294793}. 

In light of the above, one fundamental research questions is asked, 
\textbf{RQ1}: \textit{Can we learn a recommendation model that allows for higher fairness without significantly compromising recommendation accuracy?} 
And a more challenging one is, 
\textbf{RQ2}: \textit{Can we learn a single recommendation model that is able to produce optimal recommendation policies under different levels of fairness-utility trade-off so that it would be more desirable for decision-makers of e-commerce platforms to specify any preference of one objective over another based on their current business needs?}

To deal with \textbf{RQ1}, one conspicuous approach is to seek a Pareto optimal solution to guarantee optimal compromises between utility and fairness, where a Pareto efficient/optimal solution means no single objective can be further improved without hurting the others.
To find solutions with different levels of trade-off between utility and fairness  (\textbf{RQ2}), we need to generalize their Pareto frontier in the objective space, where Pareto frontier denotes a set, whose elements are all Pareto optimal.
Unfortunately, state-of-the-art approaches of fairness-aware recommendation are limited in understanding the fairness-utility trade-off.



Therefore, in this work, we aim to address the above problems and propose a fairness-aware recommendation framework using multi-objective reinforcement learning (MORL) with linear preferences, called MoFIR, which aims to learn a single parametric representation for optimal recommendation policies over the space of all possible preferences.
Technically, we first formulate the fairness-aware recommendation task as a Multi-Objective Markov Decision Process (MOMDP), with one recommendation objective, e.g., CTR, and one fairness objective, e.g., item exposure fairness (our method is able to generalize to more recommendation objectives as well as more fairness objectives).
Second, we modify classic and commonly-used RL algorithm---DDPG \cite{Silver2014DeterministicPG} by introducing conditioned networks \cite{abels2019dynamic} into it, which is a representative method to deal with multi-objective reinforcement learning.
Specially, we condition the policy network and the value network directly on the preferences by augmenting them to the feature space.
Finally, we utilize the vectorized Q-value functions together with modified loss function to update the parameters.
The contributions of this work can be summarized as follows:
\begin{itemize}
    \item We study the problem of Pareto optimal/efficient fairness-utility trade-off in recommendation and extensively explore their Pareto frontier to better satisfy real-world needs;
    \item We formulate the problem into a MOMDP and solve it through a MORL framework, MoFIR, which is optimized over the entire space of preferences in a domain, and allows the trained model to produce the optimal policy for any specified preferences;
    \item Unlike prior methods for fairness-aware recommendation, the proposed framework does not employ any relaxation for objectives in the optimization problem, hence it could achieve state-of-the-art results; 
    \item Experiments on several real-world recommendation datasets verify the superiority of our framework on both fairness measures and recommendation performance when compared with all other baselines.
\end{itemize}
\section{Related Work}

\subsection{Fairness in Recommendation}
There have been growing concerns on fairness in recommendation as recommender systems touch and influence more and more people in their daily lives.
Several recent works have found various types of bias in recommendations, such as gender and race~\cite{chen2018investigating,abdollahpouri2019unfairness}, item popularity~\cite{Zhu2018,10.1145/3437963.3441824,ge2021towards}, user feedback~\cite{fu2020fairness,10.1145/3404835.3462966,li2021user} and opinion polarity~\cite{Yao2017}. 
There are two primary paradigms adopted in recent studies on algorithmic discrimination: individual fairness and group fairness. 
Individual fairness requires that each similar individual should be treated similarly, while group fairness requires that the protected groups should be treated similarly to the advantaged group or the populations as a whole.
Our work focuses on the item popularity fairness from a group level, yet it can be used to solve multiple types of fairness simultaneously by properly defining and adding them as additional objectives.

The relevant methods related to fairness in ranking and recommendation can be roughly divided into three categories: pre-processing, in-processing and post-processing algorithms \cite{10.1145/3404835.3462807,li2021tutorial,li2021cikm}.
First of all, pre-processing methods usually aim to minimize the bias in data as bias may arise from the data source. 
This includes fairness-aware sampling methodologies in the data
collection process to cover items of all groups, or balancing methodologies to increase coverage of minority groups, or repairing methodologies to ensure label correctness, remove disparate impact \cite{10.1145/3404835.3462807}.
However, most of the time, we do not have access to the data collection process, but are given the dataset.
Secondly, in-processing methods aim at encoding fairness as part of the objective function, typically as a regularizer \cite{abdollahpouri2017controlling,beutel2019fairness}.
Finally, post-processing methods tend to modify the presentations of the results, e.g., re-ranking through linear programming \cite{li2021user,singh2018fairness,yang2021maximizing} or multi-armed bandit \cite{celis2019controlling}.
However, there is no free lunch, imposing fairness constraints to the main learning task introduces a trade-off between these objectives, which have been asserted in several studies \cite{kamani2021pareto,kearns2019ethical,NEURIPS2018_8e038477,10.5555/3294771.3294793}, e.g., \citeauthor{dutta2020there} \cite{dutta2020there} showed that because of noise on the underrepresented groups the trade-off between accuracy and equality of opportunity exists.

Unfortunately, there is very few work of fairness-aware recommendation that can be found to study the fairness-utility trade-off.
The closest one to our work is \cite{wu2021multi}, which mainly focused on the trade-off between two-sided fairness in e-commerce recommendation.
\cite{wu2021multi} used a traditional multiple gradient descent algorithm to solve multi-objective optimization problem, meaning that they need to train one network per point on the Pareto frontier, while our MoFIR generates the full Pareto frontier of solutions in a single optimization run.
Besides, the authors relaxed all their objectives to get their differentiable approximations, which, to some extent, hurt its performance, as is shown in the experiment part, Fig. \ref{fig:pareto_frontier}.


\subsection{Multi-Objective Recommendation}
Recommendation with multiple objectives is a significant but challenging problem, with the core difficulty stemming from the potential conflicts between objectives.
In most real-world recommendation systems, recommendation accuracy (e.g., CTR-oriented objectives) is the dominating factor, while some studies believed that other characteristics, such as usability, profitability, usefulness, or diversity should be considered at the same time~\cite{jambor2010optimizing,mcnee2006being,10.1145/2507157.2507226}.
When multiple objectives are concerned, it is expected to get a Pareto optimal/efficient recommendation~\cite{lin2019pareto,ribeiro2012pareto,xie2021personalized}.

The approaches on recommendation with multiple objectives to achieve Pareto efficiency can be categorized into two groups: evolutionary algorithm~\cite{zitzler2001spea2} and scalarization~\cite{lin2019pareto}.
\citeauthor{ribeiro2012pareto} \cite{ribeiro2012pareto,ribeiro2014multiobjective} jointly considered multiple trained recommendation algorithms with a Pareto-efficient manner, and conducted an evolutionary algorithm to find the appropriate parameters for weighted model combination.
Besides, \citeauthor{lin2019pareto} \cite{lin2019pareto} optimized GMV and CTR in e-commerce simultaneously based on multiple-gradient descent algorithm, which combines scalarization with Pareto-efficient SGD, and used a relaxed KKT condition. 
Our proposed method, MoFIR, belongs to scalarization, however, compared with earlier attempts in multi-objective recommendation~\cite{lin2019pareto,wu2021multi}, our method learns to adapt a single network for all the trade-off combinations of the inputted preference vectors, therefore it is able to approximate all solutions of the Pareto frontier after a single optimization run.

\subsection{RL for Recommendation}
RL-based recommenders have recently become an important and attractive topic, as it is natural to model the recommendation process as a Markov Decision Process (MDP) and use RL agents to capture the dynamics in recommendation scenarios \cite{shani2005mdp,mahmood2007learning,mahmood2009improving,zheng2018drn,xian2019reinforcement,xian2020cafe}.
Generally speaking, RL-based recommendation systems can be further classified into two categories: \textit{policy-based} \cite{Dulac-ArnoldESC15, chen2019large, chen2019top} or \textit{value-based} \cite{zhao2018recommendations, zheng2018drn,pei2019value} methods.
On one hand, policy-based methods aim to learn strategies that generate actions based on state (such as recommending items).
These methods are optimized by policy gradient, which can be deterministic approaches \cite{Dulac-ArnoldESC15, Silver2014DeterministicPG, Lillicrap2016DDPG} or stochastic approaches \cite{chen2019large, chen2019top}.
On the other hand, value-based methods aims to model the quality (e.g. Q-value) of actions so that the best action corresponds to the one with the highest Q-value.
Apart from using RL in general recommendation task, there also existed several works focusing on using RL in explainable recommendation through knowledge graphs~\cite{xian2020cafe,xian2019reinforcement}.

Currently, there are very few studies using MORL in recommendation.
\citeauthor{xie2021personalized} \cite{xie2021personalized} studied multi-objective recommendation to capture users’ objective-level preferences.
However, unlike our proposed MoFIR, which learns a single parametric representation for optimal recommendation policies, they conducted a Pareto-oriented RL to generate the personalized objective weights in scalarization for each user, which is a totally different problem formulation.

\section{Preliminary}
\subsection{Markov Decision Processes}
In reinforcement learning, agents aim at learning to act in an environment in order to maximize their cumulative reward. 
A popular model for such problems is Markov Decision Processes (MDP), which is a tuple $M = (\mathcal{S}, \mathcal{A}, \mathcal{P}, \mathcal{R}, \mu, \gamma)$, where $S$ is a set of $n$ states,  $\mathcal{A}$ is a set of $m$ actions,
$\mathcal{P}: \mathcal{S} \times \mathcal{A} \times \mathcal{S} \rightarrow [0,1]$ denotes the transition probability function, $\mathcal{R}: \mathcal{S} \times \mathcal{A} \times \mathcal{S} \rightarrow \mathbb{R}$ is the reward function, $\mu: \mathcal{S} \rightarrow [0,1]$ is the starting state distribution, and $\gamma \in [0,1)$ is the discount factor. 
We denote the set of all stationary policies by $\Pi$, where a stationary policy $\pi \in \Pi: \mathcal{S} \rightarrow P(\mathcal{A})$ is a map from states to probability distributions over actions, with $\pi(a | s)$ denoting the probability of selecting action $a$ in state $s$. 
We aim to learn a policy $\pi \in \Pi$, able to maximize a performance measure, $J(\pi)$, which is typically taken to be the infinite horizon discounted total return, 
\begin{equation}
\label{eq:discounted_return}
\small
    J(\pi) \doteq \underset{\tau \sim \pi}{\mathrm{E}}\left[\sum_{t=0}^{\infty} \gamma^{\top} R\left(s_{t}, a_{t}, s_{t+1}\right)\right],
\end{equation}
where $\tau$ denotes a trajectory, e.g., $\tau=(s_{0}, a_{0}, s_{1}, a_{1}, \dots)$, and $\tau \sim \pi$ indicates that the distribution over trajectories depends on $\pi$ :
$s_{0} \sim \mu, a_{t} \sim \pi\left(\cdot | s_{t}\right), s_{t+1} \sim P\left(\cdot | s_{t}, a_{t}\right)$.
We denote $R(\tau)$ as the discounted rewards of a trajectory, the on-policy value function as $V^{\pi}(s) \doteq$ $\mathrm{E}_{\tau \sim \pi}\left[R(\tau) | s_{0}=s\right]$, the on-policy action-value function as $Q^{\pi}(s, a) \doteq \mathrm{E}_{\tau \sim \pi}\left[R(\tau) | s_{0}=s, a_{0}=a\right]$, and the advantage function as  $A^{\pi}(s, a) \doteq Q^{\pi}(s, a)-V^{\pi}(s)$.

\subsection{Multi-Objective Markov Decision Processes}
Multi-Objective Markov Decision Processes (MOMDP) are MDPs with a vector-valued reward function $\textbf{r}_t=\textbf{R}(s_t,a_t)$, where each component of $\textbf{r}_t$ corresponds to one certain objective. 
A scalarization function \textit{f} maps the multi-objective value of a policy $\pi$  to a scalar value. 
In this work, we consider the commonly-used class of MOMDPs
with linear preference functions, e.g., $\textit{f}_{\bm{\omega}}(\textbf{R}(s,a))=\bm{\omega} \cdot \textbf{R}(s,a)$. 
It is worth noting that if $\bm{\omega}$ is fixed to a
single value, this MOMDP collapses into a standard MDP. 
An optimal solution for an MOMDP under linear \textit{f} is a convex coverage set (CCS), e.g., a set of undominated policies containing at least one optimal policy for any linear scalarization.

\subsection{Conditioned Network}
\begin{figure*}[]
\vspace{-10pt}
\centering
\mbox{
\hspace{-10pt}
\centering
    \includegraphics[scale=0.3]{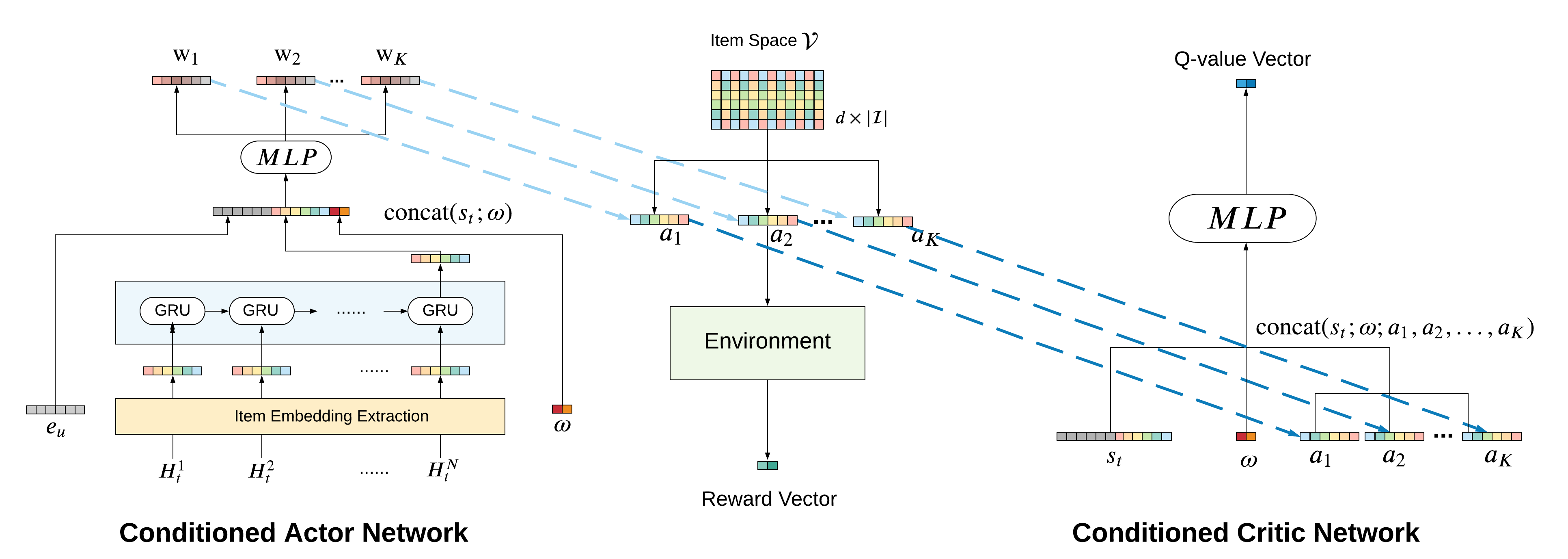}
}
\caption{The architecture of the proposed MoFIR.}
\label{fig:architecture}
\vspace{-10pt}
\end{figure*}

\citeauthor{abels2019dynamic} \cite{abels2019dynamic} studied multi-objective reinforcement learning with linear preferences and proposed a novel algorithm for learning a single Q-network that is optimized over the entire space of preferences in a domain.
The main idea is called Conditioned Network (CN), in which a Q-Network is augmented to output weight-dependent multi-objective Q-value-vectors, as is shown in the right side of Fig. \ref{fig:architecture}   (Conditioned Critic Network, where action and state representations together with weight vector are inputed to the network).
Besides, to promote quick convergence on the new weight vector’s policy and to maintain previously learned policies, the  authors updated each experience tuple in a mini-batch with respect to the current weight vector and a random previously encountered weight vector.
Specially, given a mini-batch of trajectories, they computed the loss for a given trajectory $(s_j,a_j,\textbf{r}_j,s_{j+1})$ as the sum of the loss on the active weight vector $\bm{\omega}_t$ and on $\bm{\omega}_j$ randomly sampled from the set of encountered weights.
\begin{equation}\label{eq:cn}
\begin{aligned}
&\frac{1}{2}\left[\left|\mathbf{y}_{\bm{\omega}_{t}}^{(j)}-\mathbf{Q}_{C N}\left(a_{j}, s_{j} ; \bm{\omega}_t\right)\right|+\left|\mathbf{y}_{\bm{\omega}_{j}}^{(j)}-\mathbf{Q}_{C N}\left(a_{j}, s_{j} ; \bm{\omega}_j\right)\right|\right]
\end{aligned}
\end{equation}

\begin{equation}
\begin{aligned}
\textbf{y}_{\bm{\omega}}^{(j)}=\textbf{r}_{j}+\gamma \textbf{Q}_{C N}^{-}\left(\underset{a \in A}{\operatorname{argmax}} \textbf{Q}_{C N}\left(a, s_{j+1} ;\bm{\omega}\right) \cdot \bm{\omega}, s_{j+1} ; \bm{\omega}\right)
\end{aligned}
\end{equation}
where $\textbf{Q}_{C N}\left(a, s ;\bm{\omega}\right)$ is the network’s Q-value-vector for action $a$ in state $s$ and with weight vector $\bm{\omega}$.
They claimed that training the same sample on two different weight vectors has the added advantage of forcing the network to identify that different weight vectors can have different Q-values for the same state.
A more comprehensive review of MOMDPs and CN can be seen in \cite{abels2019dynamic}.

In the original paper, the authors only proposed an algorithm based on Double DQN with discrete action space, which is not suitable for recommendation scenarios as the action space of recommendation is very large.
Therefore, we modify the traditional DDPG \cite{Silver2014DeterministicPG} by introducing  conditioned network into its policy network as well as critic network, and more importantly, we modify the original loss functions for both of them. 
We choose DDPG as it is a commonly adopted methods in RL, while our modification can be generalized to other reinforcement learning methods, such as trust region ploicy optimization.
More details about our modification will be introduced in Section \ref{sec:framework}.

\section{Problem Formulation}

\subsection{\textbf{MOMDP for Recommendation}} 
The recommendation agent will take the feature representation of the current user and item candidates $\mathcal{I}$ as input, and generate a list of items $L\in\mathcal{I}^K$ to recommend, where $K \geq 1$ after a user sends a request to it at timestamp $t \in$ ($t_1$, $t_2$, $t_3$, $t_4$, $t_5$, $\dots$).
User $u$ who has received the list of recommended items $L$ will give feedback $B$ via clicking on this set of items, which can be used to measure the recommendation performance.
Besides, based on the recommendation results, we will acquire the total number of exposure for each item group $G$, which can later be used to measure fairness.
Thus, the state $s$ can be represented by user features (e.g., user's recent click history), action $a$ is represented by items in $L$, reward $\textbf{r}$ is the immediate reward vector after taking action $a$, with each component of $\textbf{r}$ corresponds to one certain objective (e.g., whether user clicks on an item in $L$ for utility objective or whether an item comes from predefined disadvantageous group for fairness  objective).
The problem formulation is formally presented as follows:
\begin{itemize}
    \item \textbf{State $\mathcal{S}$:} A state $s_t$ is the representation of user’s most recent positive interaction history $H_t$ with the recommendation system, together with his/her demographic information (if exists).
    
    \item \textbf{Action $\mathcal{A}$:} An action $a_t\ = \{ a_t^{1},\ \dots,\ a_t^{K}\}$ is a recommendation list with $K$ items to a user $u$ at time $t$ with current state $s_t$.
    
    \item \textbf{Vector Reward Function $\bm{r}$:} A vector-valued reward function $\textbf{r}_t=\textbf{R}(s_t,a_t)$, where each component of $\textbf{r}_t$ corresponds to one certain objective. In this work, the reward vector includes two elements: utility objective and fairness objective. The details of the definition of our task-specific objectives will be introduced in the following section.
    
    \item \textbf{Scalarization function}  $\textit{f}$ \textbf{:} In this paper, we consider the class of MOMDPs with linear preferences functions $\textit{f}$, which is a commonly-used scalarization function.
    Under this setting, each objective is given a weight $\omega_i$, such that the scalarization function becomes $\textit{f}_{\bm{\omega}}(\textbf{R})=\bm{\omega} \cdot \textbf{R}$, where each $\omega_i \in [0,1]$  and $\sum_{i} \omega_i = 1$. 
    
    \item \textbf{Discount rate $\mathcal{\gamma}$:} $\mathcal{\gamma} \in [0,1]$ is a discount factor measuring the present value of long-term rewards.
    
\end{itemize}

We aim to learn a policy $\pi$, mapping from states to actions, to generate recommendations that achieve the Pareto efficient trade-off  between fairness and utility.

\subsection{Multi-Objectives in Fair Recommendation}
The reward vector is designed to measure the recommendation system’s gain regarding utility and fairness.
While our method is capable of dealing with multiple objectives simultaneously, for simplicity we deliberately select click through rate and item (group) exposure fairness as our two objectives recommendation utility and item exposure fairness respectively.

\subsubsection{Utility Objective}
On one hand, given the recommendation based on the action $a_t$ and the user state $s_t$, the user will provide feedback, e.g. click or purchase, etc. 
The recommender receives immediate reward $R_u(s_t, a_t)$ according to the user’s positive feedback. We also normalize the reward value by dividing $K$, which is the length of the recommendation list.

\begin{equation} \label{eq:reward}
\centering
\small
\begin{aligned}
    R_u(s_{t},a_{t}, s_{t+1}) &= \frac{\sum_{l=1}^{K} \vmathbb{1} (a_{t}^l \text{ gets positive feedback})}{K}\\
\end{aligned}
\end{equation}

\subsubsection{Fairness  Objective}
On the other hand, based on the recommendation list $a_t$, the total number of exposure of each item group will be counted and used to measure exposure fairness.
Here, we calculate the ratio of items from sensitive group to the total number of recommended items, and use a hinge loss with margin $\beta$ to punish the abuse of fairness. Usually, we set $\beta$ to be the ratio of the number of items in sensitive group to the total number of items.

\begin{equation} \label{eq:cost}
\small
\centering
\begin{aligned}
    R_f(s_{t},a_{t}, s_{t+1}) &=  \max \left( \frac{\sum_{l=1}^{K} \vmathbb{1}(a_{t}^l\ is\ in\ sensitive\ group)}{K}, \beta  \right)
\end{aligned}
\end{equation}


\section{Proposed Framework}\label{sec:framework}

\begin{algorithm}[h]
\small
    \textbf{Input:} \\
    A preference sampling distribution $D_\omega$;\\
    A  multi-objective critic network $\mu$ parameterized by $\theta^\mu$;\\
    An actor network  $\pi$ parameterized by $\theta^\pi$;\\
    Pre-trained user embeddings $\mathcal{U}$ and item embeddings $\mathcal{V}$.\\
    
    \textbf{Output:} \\
    Parameters $\theta^\pi$, $\theta^\mu$ of the actor network and critic network.\\
    \textbf{Initialization:}\\
    Randomly initialize $\theta^\pi$ and $\theta^\mu$;\\
    Initialize target network $\mu^{\prime}$ and $\pi^{\prime}$ with weights ${\theta^\pi}^{\prime} \leftarrow \theta^\pi, {\theta^\mu}^{\prime} \leftarrow \theta^\mu$;\\
    Initialize replay buffer $D$.\\
    

    \For{$\text{Episode}\ =\ 1\ ...\ M$}{
        Initialize user state $s_0$ from log data;\\
        Sample a linear preference $\bm{\omega}_{0} \sim D_\omega$;\\
        \For{$t\ =\ 1\ ...\ T$}{
        Observe current state, represent it as $s_t$ based on Eq. \eqref{eq:state_rep};\\
        Select an action $a_t\ \in \mathcal{I}^K$ using actor network $\pi$ based on Eq. \eqref{eq:action}\\
        Calculate utility reward and fairness reward and get the multi-objective reward vector $\textbf{r}_t$ according to environment feedback based on Eq. \eqref{eq:reward} and Eq. \eqref{eq:cost};\\
        Update $s_{t+1}$ based on Eq. \eqref{eq:state_rep};\\
        Store transition $(s_t,a_t,\textbf{r}_t,s_{t+1})$ in $D$.
        }
        
        \If{update}{
        Sample minibatch of $\mathcal{N}$ trajectories $\mathcal{T}$ from $D$;\\
        Sample $\mathcal{N}_\omega$ preferences $\textit{W}  = \{\bm{\omega}_1,\bm{\omega}_2,\dots,\bm{\omega}_{N_\omega}\} \sim D_\omega$;\\
        Append $\bm{\omega}_0$ to $\textit{W}$;\\
        Select an action $a^{\prime} \in \mathcal{I}^K$ using actor target network $\pi^{\prime}$;\\
        Set $\textbf{y}=\textbf{r}+\gamma \textbf{Q}^{\prime}\left(s^{\prime}, a^{\prime},\bm{\omega}; {\theta^\mu}^{\prime}\right), \bm{\omega} \in \textit{W}$\\
        Update critic by minimizing $ \|\textbf{y}-\textbf{Q}\left(s, a, \bm{\omega}; {\theta^\mu} \right) \|_{2}^{2}$ according to:
        $$
        \nabla_{{\theta^\mu}} \mathcal{L} \approx \frac{1}{\mathcal{N} \mathcal{N}_\omega}\left[\left(\textbf{y}-\textbf{Q} \left(s, a, \bm{\omega} ; {\theta^\mu} \right)\right)^{T} \nabla_{\theta^\mu} \textbf{Q}\left(s, a, \bm{\omega} ; {\theta^\mu} \right)\right]
        $$\\
        Update the actor using the sampled policy gradient:
        $$
        \nabla_{\theta^\pi} \pi \approx \frac{1}{\mathcal{N} \mathcal{N}_\omega} \sum_{i} \bm{\omega}^{T} \nabla_{a}\textbf{Q}\left(s, a, \bm{\omega}; \theta^\mu \right) \nabla_{\theta^\pi} \pi(s,\bm{\omega})
        $$\\
        Update the critic target networks:
        $$
        {\theta^\mu}^{\prime} \leftarrow \tau \theta^\mu+(1-\tau) {\theta^\mu}^{\prime}
        $$\\
        Update the actor target networks:
        $$
        {\theta^\pi}^{\prime} \leftarrow \tau \theta^\pi+(1-\tau) {\theta^\pi}^{\prime}
        $$
        
        }
    }
    \caption{Multi-Objective DDPG Algorithm}
    \label{alg:MoFIR}
\end{algorithm}

\subsection{The Conditioned Actor}
The conditioned actor is almost the same as traditional actor except that we condition the predictions of the policy network to the preference vectors.
Practically, we concatenate the state representation $s_t$ with the vector $\bm{\omega}$ and train a neural network on this joint feature space, which is depicted in Fig. \ref{fig:architecture} (Conditioned  Actor Network).
The conditioned actor $\pi$ parameterized by $\theta^\pi$ serves as a stochastic policy that samples an action $a_t\in\mathcal{I}^K$ given the current state $s_t\in\mathbb{R}^m$ of a user and the  preference vector $\bm{\omega}$.

First of all, we define $s_t$ as the concatenation of the user embedding $\mathbf{e}_u\in\mathbb{R}^d$ and their recent history embedding $\mathbf{h}_u$:
\begin{equation}\label{eq:state_rep}
s_t = [\mathbf{e}_u; \mathbf{h}_u],
\end{equation}
where the recent history embedding $\mathbf{h}_u=\mathrm{GRU}(H_t)$ is acquired by encoding $N$ item embeddings via Gated Recurrent Units (GRU) \cite{Cho2014gru}, and $H_t = \{H_t^1, H_t^2, \dots, H_t^N\}$ denotes the most recent $N$ items from user $u$'s interaction history.
We define the user's recent history is organized as a queue with fixed length, and update it only if the recommended item $a_t^l \in a_t$ receives a positive feedback, which ensures that the state can always represent the user's most recent interests.
\begin{equation}
\small
    H_{t+1}=\left\{
    \begin{array}{cc}
         \{H_t^2,\ \dots,\ H_t^N,\ a_t^l\} & a_t^l \text{ gets positive feedback} \\
         H_t & \text{Otherwise}
    \end{array}
    \right.
\end{equation}

Secondly, we assume that the probability of actions conditioned on states and preferences follows a continuous high-dimensional Gaussian distribution.
We also assume it has mean $\mu\in\mathbb{R}^{Kd}$ and covariance matrix $\Sigma\in\mathbb{R}^{Kd\times Kd}$ (only elements at diagonal are non-zeros and there are actually $Kd$ parameters).
In order to achieve better representation ability, we approximate the distribution via a deep neural network, which maps the encoded state $s_t$ and preferences $\bm{\omega}$ to $\mu$ and $\Sigma$.
Specifically, we adopt a Multi Layer Perceptron (MLP) with tanh($\cdot$) as the non-linear activation function,  
\begin{equation}
    (\mu,\Sigma)=\mathrm{MLP}(s_t,\bm{\omega}).
\end{equation}
Once received $\mu$ and $\Sigma$, we sample a vector from the acquired Gaussian distribution $\mathcal{N}(\mu,\Sigma)$ and convert it into a proposal matrix $W\sim \mathcal{N}(\mu,\Sigma)\in\mathbb{R}^{K\times d}$, whose $k$-th row, denoted by $W_k\in\mathbb{R}^d$, represents an ``ideal'' embedding of a virtual item.

Finally, the probability matrix $P\in\mathbb{R}^{K\times |\mathcal{I}|}$ of selecting the $k$-th candidate item is given by
$
\small
    P_k = \mathrm{softmax}(W_k \mathcal{V}^\top),~ k=1,\ldots,K,
$
where $\mathcal{V}\in\mathbb{R}^{|\mathcal{I}|\times d}$ is the embedding matrix of all candidate items.
This is equivalent to using dot product to determine similarity between $W_k$ and any item.
As the result of taking the action at step $t$, the actor recommends the $k$-th item as follows:
\begin{equation} \label{eq:action}
\small
    a_t^k = \argmax_{i\in \{1,\dots,|\mathcal{I}|\}} P_{k,i},~ \forall k=1,\ldots,K,
\end{equation}
where $P_{k,i}$ denotes the probability of taking the $i$-th item at rank $k$.

\subsection{The Conditioned Critic}
The conditioned critic $\mu$ also differs from the traditional critic in that we concatenate the state representation $s_t$ with the vector $\bm{\omega}$ as well as the embedding of action $a_t$, and require the output to be a Q-value-vector with the size equal to the number of  objectives, which is depicted in Fig. \ref{fig:architecture} (Conditioned Critic Network).
The conditioned critic $\mu$ is parameterized with $\theta^\mu$ and is constructed to approximate the true state-action value vector function $\textbf{Q}^\pi(s_t,a_t,\bm{\omega})$ and is used in the optimization of the actor.
Following Eq. \ref{eq:cn} introduced in conditioned network \cite{abels2019dynamic}, the conditioned critic network is updated according to temporal-difference learning that minimizes the following loss function:
\begin{equation} \label{eq:value_update}
\small
    \mathcal{L}(\theta^\mu) = \mathbb{E}_{s, a, \boldsymbol{\omega}}\left[\|\textbf{y}_t-\textbf{Q}_t\left(s, a, \bm{\omega}; \theta^\mu \right) \|_{2}^{2}\right]
\end{equation}
where $\textbf{y}_t = \textbf{r}_t + \gamma \textbf{Q}_{\omega}(s_{t+1},a_{t+1},\bm{\omega};\theta^\mu)$.

\subsection{Parameters Training Procedure of MoFIR}
We present the detailed training procedure of our proposed model, MoFIR, in Algorithm \ref{alg:MoFIR} and the model architecture in Fig. \ref{fig:architecture}. 
As mentioned before, we modify traditional single-objective DDPG into multi-objective DDPG by introducing the conditioned networks to both its actor network and critic network.
In each episode, there are two phases --- the trajectory generation phase (line 15-20) and model updating phase (line 22-32).
In the trajectory generation phase, we sample one linear preference $\bm{\omega}_0$ and fix it to generate user-item interaction trajectories.
Then in the model updating phase,  we sample another $\mathcal{N}_\omega$ preferences together with $\bm{\omega}_0$ to update the conditioned actor network and the conditioned critic network.
Here, we do not follow the original setting in \cite{abels2019dynamic}, which only uses one more random sampled preference vector, 
as \citeauthor{yang2019generalized} \cite{yang2019generalized} observed that increasing the number of sampled preference vectors can further improve the coverage ratio of RL agent and diminish the adaptation error in their experiments.

\begin{table}[h]
\caption{Basic statistics of the experimental datasets.}
\label{tab:dataset}
\centering
\setlength{\tabcolsep}{5pt}
\begin{adjustbox}{max width=\linewidth}
\begin{tabular}
    {lcccccccc} \toprule
    Dataset & \#users & \#items & \#act./user & \#act./item & \#act. & density \\\midrule
    Movielens100K & 943 & 1682 & 106 & 59.45 & 100,000 & 6.305\%\\
    Ciao & 2248 & 16861 & 16 & 2 & 36065 & 0.095\%\\
    Etsy & 1030 & 945 & 47 & 51 &  48080 & 4.940\%\\\bottomrule
\end{tabular}
\end{adjustbox}
\end{table}

\section{Experiments}
In this section, we first introduce the datasets, the comparison baselines, then discuss and analyse the experimental results.

\subsection{Dataset Description}
To evaluate the models under different data scales, data sparsity and application scenarios, we perform experiments on three real-world datasets.
Some basic statistics of the experimental datasets are shown in Table \ref{tab:dataset}.

\begin{itemize}[leftmargin=*]
    \item \textbf{Movielens:} We choose Movielens100K \footnote{\url{https://grouplens.org/datasets/Movielens/}}, which includes about one hundred thousand user transactions, respectively (user id, item id, rating, timestamp, etc.).
    
    \item \textbf{Ciao:} Ciao was collected  by \citeauthor{10.1145/2339530.2339574} \cite{10.1145/2339530.2339574} from a popular product review site, Epinions, in the month of May, 2011\footnote{https://www.cse.msu.edu/~tangjili/datasetcode/truststudy.htm}. For each user, they  collected user profiles, user ratings and user trust relations. For each rating, they collected the product name and its category, the rating score, the time point when the rating is created, and the helpfulness of this rating.
    
    \item \textbf{Etsy:} We collect a few weeks of user-item interaction data on a famous e-commerce platform, Etsy. For each record, we collect user id, item id and timestamp. Since the original data is sparse, we filter out users and items with fewer than twenty interactions.
\end{itemize}

For each dataset, we first sort the records of each user based on the timestamp, and then split the records into training and testing sets chronologically by 4:1. 
The last item in the training set of each user is put into the validation set.
Since we focus on item exposure fairness, we need to split items into two groups $G_0$ and $G_1$ based on item popularity.
It would be desirable if we have the item impression/listing information and use it to group items, however, since Movielens and Ciao are public dataset and only have interaction data, we use the number of interaction to group items in them.
Specifically, for Movielens and Ciao, the top 20\% items in terms of number of interactions belong to the popular group $G_0$, and the remaining 80\% belong to the long-tail group $G_1$, while for Etsy data, we additionally collect the listing impressions per month for each item and group items based on this.

Moreover, for RL-based methods, we set the initial state for each user during training as the first five clicked items in the training set, and the initial state during testing as the last five clicked items in the training set.
We also set the RL agent recommend ten items to a user each time.

\begin{table*}[]
\caption{Summary of the performance on three datasets. 
We evaluate for ranking ($Recall$, $F1$ and $NDCG$, in percentage (\%) values, \% symbol is omitted in the table for clarity) and fairness ($KL\ Divergence$ and $Popularity\ Rate$, also in \% values), while $K$ is the length of recommendation list. 
Bold scores are used when MoFIR is the best, while underlined scores indicate the strongest baselines.
When MoFIR is the best, its improvements against the best baseline are significant at p < 0.01.}
\centering
\begin{adjustbox}{max width=\linewidth}
\setlength{\tabcolsep}{7pt}
\begin{tabular}
    {m{1.53cm} ccc ccc ccc ccc ccc} \toprule
    \multirow{2}{*}{Methods} 
    & \multicolumn{3}{c}{Recall (\%) $\uparrow$} 
    & \multicolumn{3}{c}{F1 (\%) $\uparrow$} 
    & \multicolumn{3}{c}{NDCG (\%) $\uparrow$} 
    & \multicolumn{3}{c}{KL (\%) $\downarrow$} 
    & \multicolumn{3}{c}{Popularity Rate (\%) $\downarrow$}\\\cmidrule(lr){2-4} \cmidrule(lr){5-7} \cmidrule(lr){8-10} \cmidrule(lr){11-13} \cmidrule(lr){14-16}
 & K=5 & K=10 & K=20 & K=5 & K=10 & K=20 & K=5 & K=10 & K=20 & K=5 & K=10 & K=20 & K=5 & K=10 & K=20 \\\midrule 

\multicolumn{16}{c}{Movielens-100K} \\\midrule
MF  &  1.422 & 2.713 & 5.228 & 2.019 & 3.016 & 4.127 & 3.561 & 3.830 & 4.705 & 229.124 & 224.390 & 215.772 & 99.745 & 99.258 & 98.224 \\
BPR-MF  & 1.304 & 3.539 & 8.093 & 1.824 & 3.592 & 5.409 & 3.025 & 3.946 & 5.787 & 230.531 & 230.531 & 229.464 & 99.873 & 99.873 & 99.777 \\
NGCF  & 1.995 & 3.831 & 6.983 & 2.846 & 4.267 & 5.383 & 5.319 & 5.660 & 6.510 & 232.193 & 232.193 & 232.193 & 100.000 & 100.000 & 100.000 \\
LIRD  & \underline{2.798} & \underline{6.586} & \underline{13.711} & \underline{3.198} & \underline{4.850} & \underline{5.855} &\underline{ 4.583} & \underline{6.217} & \underline{8.840} & 209.845 & 193.918 & 176.644 & 97.434 & 95.058 & 92.121 \\
\midrule

MF-FOE & 1.164 & 2.247 & 4.179 & 1.739 & 2.730 & 3.794 & 3.520 & 3.796 & 4.367 & 181.000 & 175.355 & 170.444 & 92.895 & 91.888 & 90.981 \\
BPR-FOE & 0.974 & 2.053 & 4.404 & 1.496 & 2.568 & 3.933 & 3.127 & 3.514 & 4.332 & 176.938 & 172.465 & 168.952 & 92.174 & 91.357 & 90.700 \\
NGCF-FOE & 1.193 & 1.987 & 4.251 & 1.759 & 2.398 & 3.698 & 4.033 & 3.897 & 4.633 & 232.193 & 232.193 & 232.193 & 100.000 & 100.000 & 100.000 \\\midrule

MF-MFR & 1.546 & 2.807 & 5.422 & 2.019 & 3.016 & 4.127 & 3.276 & 3.613 & 4.571 & \underline{100.590} & \underline{96.620 }& \underline{85.420 }& \underline{74.867 }& \underline{73.743 }& \underline{70.419 }\\
BPR-MFR & 1.418 & 2.811 & 6.155 & 2.019 & 3.016 & 4.127 & 3.522 & 3.822 & 5.047 & 165.897 & 155.339 & 137.663 & 90.117 & 88.017 & 84.205 \\
NGCF-MFR & 1.456 & 2.900 & 6.570 & 2.846 & 4.267 & 5.383 & 3.041 & 3.472 & 4.928 & 212.497 & 202.306 & 185.518 & 97.794 & 96.352 & 93.674 \\\midrule

MoFIR-1.0 & \textbf{6.580} & \textbf{12.753} & \textbf{22.843} & \textbf{5.658} & \textbf{7.178} & \textbf{7.858} & \textbf{8.026} & \textbf{10.848} & \textbf{14.683} & 232.193 & 232.193 & 232.193 & 100.000 & 100.000 & 100.000 \\
MoFIR-0.5 & 4.679 & 9.520 & 19.918 & 4.438 & 5.808 & 7.421 & 6.633 & 9.300 & 14.370 & 173.672 & 170.303 & 162.386 & 91.580 & 90.954 & 89.433 \\
MoFIR-0.1  & 0.323 & 0.781 & 1.550 & 0.521 & 1.008 & 1.483 & 1.251 & 1.404 & 1.598 & \textbf{0.795} & \textbf{0.608} & \textbf{0.306} & \textbf{24.305} & \textbf{23.754} & \textbf{22.646} \\
\midrule

\multicolumn{16}{c}{Ciao} \\\midrule
MF  &  0.518 & 1.938 & 3.100 & 0.395 & 0.687 & 0.599 & 0.408 & 0.924 & 1.264 & 81.154 & 65.458 & 47.848 & 69.088 & 63.835 & 57.098 \\
BPR-MF  & 1.087 & 2.204 & 4.607 & 0.677 & 0.770 & 0.858 & 0.776 & 1.181 & 1.900 & 119.307 & 100.884 & 82.717 & 79.826 & 74.949 & 69.580 \\
NGCF & \underline{1.721} & \underline{2.816} & \underline{4.380} & \underline{1.056} & 0.958 & 0.783 & \underline{1.670} & 2.027 & 2.450 & 142.025 & 96.789 & 59.561 & 85.181 & 73.792 & 61.693 \\
LIRD & 0.766 & 2.448 & 3.599 & 0.554 & \underline{1.082} & \underline{0.921} & 1.393 & \underline{2.638} & \underline{3.277} & 65.744 & 105.507 & 64.888 & 63.936 & 76.223 & 63.632 \\\midrule

MF-FOE & 0.685 & 1.208 & 1.914 & 0.458 & 0.474 & 0.396 & 0.475 & 0.669 & 0.864 & 19.720 & 11.167 & 7.622 & 43.068 & 37.033 & 33.915 \\
BPR-FOE & 1.442 & 2.111 & 3.693 & 0.812 & 0.663 & 0.731 & 0.934 & 1.154 & 1.657 & 55.999 & 46.858 & 40.626 & 60.347 & 56.686 & 53.987 \\
NGCF-FOE & 1.234 & 1.907 & 2.903 & 0.651 & 0.583 & 0.566 & 0.937 & 1.156 & 1.477 & 79.313 & 74.038 & 71.335 & 43.357 & 34.226 & 30.391 \\\midrule

MF-MFR & 0.307 & 0.619 & 1.281 & 0.395 & 0.687 & 0.599 & 0.237 & 0.345 & 0.535 & \underline{0.185 }& \underline{0.096 }& \underline{0.068 }& \underline{18.003} & \underline{18.553} & \underline{18.784} \\
BPR-MFR & 1.146 & 1.962 & 2.667 & 0.395 & 0.687 & 0.599 & 1.011 & 1.314 & 1.534 & 4.303 & 2.540 & 1.454 & 30.304 & 27.829 & 25.868 \\
NGCF-MFR & 1.284 & 2.131 & 4.033 & 1.056 & 0.958 & 0.783 & 1.014 & 1.342 & 1.901 & 37.133 & 20.302 & 10.515 & 52.388 & 43.430 & 36.498 \\\midrule

MoFIR-1.0  &  \textbf{2.162} & \textbf{3.867} & \textbf{5.866} & \textbf{1.626} & \textbf{1.513} & \textbf{1.323} & \textbf{4.000} & \textbf{4.764} & \textbf{5.813} & 181.742 & 156.545 & 123.213 & 93.025 & 88.263 & 80.796 \\
MoFIR-0.5  & 1.254 & 2.665 & 4.122 & 0.845 & 0.971 & 0.879 & 2.031 & 2.724 & 3.490 & 19.077 & 12.750 & 8.032 & 42.663 & 38.278 & 34.305 \\
MoFIR-0.1 & 0.892 & 1.610 & 2.338 & 0.557 & 0.532 & 0.445 & 1.054 & 1.311 & 1.576 & \textbf{0.054 } & \textbf{0.010 } & 0.484 & 21.100 & 19.522 & \textbf{16.795} \\\midrule

\multicolumn{16}{c}{Etsy} \\\midrule
MF  &  2.693 & 5.581 & 10.348 & 2.917 & 4.176 & 4.912 & 3.438 & 4.671 & 6.681 & 190.410 & 190.173 & 186.243 & 94.491 & 94.452 & 93.797 \\
BPR-MF  & 3.113 & 5.850 & 11.704 & 3.309 & 4.320 & 5.385 & 3.700 & 4.880 & 7.341 & 179.815 & 176.447 & 169.740 & 92.687 & 92.085 & 90.849 \\
NGCF & 3.414 & 6.026 & 11.746 & 3.674 & \underline{4.498} & \underline{5.406} & 4.180 & 5.238 & 7.610 & 194.985 & 185.756 & 175.403 & 95.228 & 93.715 & 91.896 \\
LIRD & \underline{7.163} & \underline{12.176} & \underline{24.056} & \underline{4.158} & 4.493 & 4.967 & \underline{6.587} & \underline{9.289} & \underline{13.833} & 212.890 & 197.336 & 166.047 & 97.847 & 95.597 & 90.145 \\\midrule

MF-FOE & 1.382 & 2.436 & 4.515 & 1.641 & 2.160 & 2.704 & 2.111 & 2.482 & 3.318 & \underline{42.682} & \underline{29.960} & 22.502 & \underline{54.898} & \underline{48.865} & 44.758 \\
BPR-FOE & 1.503 & 2.808 & 5.513 & 1.802 & 2.468 & 3.132 & 2.328 & 2.783 & 3.844 & 43.394 & 30.734 & 23.390 & 55.209 & 49.263 & 45.276 \\
NGCF-FOE & 1.958 & 3.135 & 5.478 & 2.227 & 2.593 & 2.923 & 2.705 & 3.106 & 4.024 & 47.548 & 30.829 & \underline{21.678} & 56.974 & 49.311 & \underline{44.268} \\\midrule

MF-MFR & 2.482 & 5.150 & 10.279 & 2.917 & 4.176 & 4.912 & 3.265 & 4.504 & 6.671 & 173.889 & 158.030 & 134.564 & 91.620 & 88.565 & 83.497 \\
BPR-MFR & 2.510 & 4.849 & 9.711 & 2.917 & 4.176 & 4.912 & 3.144 & 4.110 & 6.206 & 136.319 & 120.661 & 94.153 & 83.899 & 80.165 & 73.031 \\
NGCF-MFR & 2.325 & 4.146 & 7.946 & 3.558 & 4.820 & 5.616 & 2.994 & 3.636 & 5.210 & 104.348 & 91.557 & 74.402 & 75.907 & 72.270 & 66.901 \\\midrule

MoFIR-1.0  &  6.690 & \textbf{13.87}1 & \textbf{24.72}8 & \textbf{4.833} & \textbf{5.932} & \textbf{6.238} & \textbf{9.183} & \textbf{13.629} & \textbf{19.822} & 139.319 & 134.627 & 129.318 & 84.578 & 83.511 & 82.270 \\
MoFIR-0.5  & 5.333 & 10.342 & 19.383 & 3.460 & 3.979 & 4.218 & 4.626 & 6.614 & 9.798 & 70.154 & 67.961 & 64.956 & 65.470 & 64.714 & 63.657 \\
MoFIR-0.1 & 1.340 & 2.966 & 5.864 & 1.151 & 1.641 & 1.895 & 1.778 & 2.839 & 4.425 & \textbf{0.569 }& \textbf{0.545 }& \textbf{0.396 }& \textbf{23.628} & \textbf{23.550} & \textbf{23.016} \\\bottomrule
\end{tabular}\label{tab:result}
\end{adjustbox}
\vspace{-10pt}
\end{table*}

\begin{figure*}[t]
\mbox{
\hspace{-15pt}
\centering
    \subfigure[NDCG vs Long-tail Rate on ML100K]{\label{fig:ml100k_ndcg_pop}
        \includegraphics[width=0.28\textwidth]{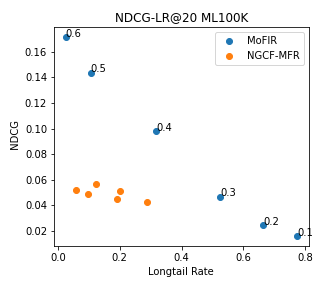}}
    \hspace{15pt}
    \subfigure[NDCG vs Long-tail Rate on Ciao]{\label{fig:ml100k_ndcg_pop}
        \includegraphics[width=0.28\textwidth]{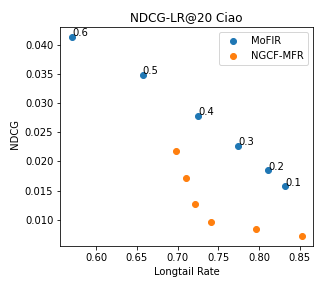}}
    \hspace{15pt}
    \subfigure[NDCG vs Long-tail Rate on Etsy]{\label{fig:ml100k_ndcg_pop}
        \includegraphics[width=0.28\textwidth]{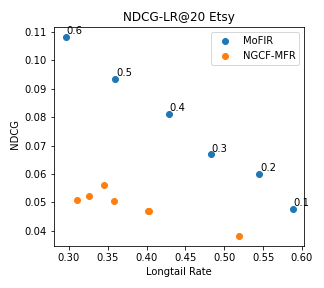}}
}
\caption{Approximate Pareto frontier in three datasets generated by MoFIR and NGCF-MFR, where $x$-axis represents the $Longtail\ Rate@20$ ($Longtail\ Rate$ equals to one minus $Popularity\ Rate$) and $y$-axis represents the value of $NDCG@20$.}
\label{fig:pareto_frontier}
\end{figure*}

\subsection{Experimental Setup}\label{sec:experimental_setup}
\subsubsection{\textbf{Baselines:}}
We compare our proposed method with the following baselines, including both traditional and reinforcement learning  based recommendation models. 

\begin{itemize}[leftmargin=*]
\item {\bf MF}: Collaborative Filtering based on matrix factorization \cite{koren2009matrix} is a representative method for rating prediction. 
However, since not all datasets contain rating scores, we turn the rating prediction task into ranking prediction.
Specifically, the user and item interaction vectors are considered as the representation vector for each user and item.

\item {\bf BPR-MF}: Bayesian Personalized Ranking \cite{bpr} is one of the most widely used ranking methods for top-K recommendation, which models recommendation as a pair-wise ranking problem.
In the implementation, we conduct balanced negative sampling on non-purchased items for model learning.

\item {\bf NGCF}: Neural Graph Collaborative Filtering \cite{wang2019neural} is a neural network-based recommendation algorithm, which integrates the user-item interactions into the embedding learning process and exploits the graph structure by propagating embeddings on it to model the high-order connectivity.

\item {\bf LIRD}: The original paper for List-wise recommendation based on deep reinforcement learning (LIRD) \cite{DBLP:journals/corr/abs-1801-00209} utilized the concatenation of item embeddings to represent the user state, and the actor will provide a list of K items as an action. 
\end{itemize}

We also include two state-of-the-art fairness frameworks to show the fairness performance of our proposed method.
\begin{itemize}[leftmargin=*]
    \item \textbf{FOE}: Fairness of Exposure in Ranking (FOE) \cite{singh2018fairness} is a type of post-processing algorithm incorporating a standard linear program and the Birkhoff-von Neumann decomposition.
    It is originally designed for searching problems, so we follow the same modification method mentioned in \cite{10.1145/3437963.3441824, wu2021multi}, and use ranking prediction model such as MF, BPR, and NGCF as the base ranker, where the raw utility is given by the predicted probability of user $i$ clicking item $j$. 
    In our experiment, we have \textbf{MF-FOE}, \textbf{BPR-FOE} and \textbf{NGCF-FOE} as our fairness baselines. Since FOE assumes independence of items in the list, it cannot be applied to LIRD, which is a sequential model and the order in its recommendation makes a difference.
    
    \item \textbf{MFR}: Multi-FR (MFR) \cite{wu2021multi} is a generic fairness-aware recommendation framework with multi-objective optimization, which jointly optimizes fairness and utility for two-sided recommendation. In our experiment, we only choose its item popularity fairness. We also modify it as the original fairness considers position bias as well, which is not the same setting as ours.
    Finally, we have \textbf{MF-MFR}, \textbf{BPR-MFR} and \textbf{NGCF-MFR}. For same reason as FOE, we do not include LIRD as well.
\end{itemize}

We implement MF, BPR-MF, NGCF, MF-FOE, BPR-FOE, NGCF-FOE, MF-MFR BPR-MFR and NGCF-MFR using \textit{Pytorch} with Adam optimizer.
For all of them, we consider latent dimensions $d$ from \{16, 32, 64, 128, 256\}, learning rate $lr$ from \{1e-1, 5e-2, 1e-2, \dots, 5e-4, 1e-4\}, and the L2 penalty is chosen from \{0.01, 0.1, 1\}. 
We tune the hyper-parameters using the validation set and terminate training when the performance on the validation set does not change within 5 epochs.
Further, since the FOE-based methods needs to solve a linear programming with size $|\mathcal{I}|\times|\mathcal{I}|$for each consumer, which brings huge computational costs, we rerank the top-200 items from the base model then select the new top-K (K<100) as the final recommendation.
Similarly, we implement \textbf{MoFIR} with $Pytorch$. 
We first perform basic MF to pretrain 16-dimensional user and item embeddings, and fix them through training and test.
We set $|H_t|=5$, and use two GRU layers to get the state representation $s_t$. 
For the actor network and the critic network, we use two hidden layer MLP with tanh($\cdot$) as activation function.
Finally, we fine-tune MoFIR's hyper-parameters on our validation set.
In order to examine the trade-off between performance and fairness, we use different level of preference vectors in test.
Since MoFIR is able to approximate all possible solutions of the Pareto frontier, we simply input different preference vetors $\bm{\omega}$ into the trained model to get variants of MoFIR and denote the resulting alternatives as \textbf{MoFIR-1.0}, \textbf{MoFIR-0.5}, and \textbf{MoFIR-0.1}, where the scalar is the weight on the recommendation utility objective.

\vspace{-5pt}
\subsubsection{\textbf{Evaluation Metrics:}} 
We select several most commonly used top-K ranking metrics to evaluate each model's recommendation performance, including \textbf{Recall}, \textbf{F1 Score}, and \textbf{NDCG}.
For fairness evaluation, we define \textbf{Popularity Rate}, which simply refers to the ratio of the number of popular items in the recommendation list to the total number of items in the list.
We also employ \textbf{KL-divergence} (KL) to compute the expectation of the difference between protected group membership at top-K vs. in the over-all population, where $d_{K L}\left(D_{1}|| D_{2}\right)=\sum_{j} D_{1}(j) \ln \frac{D_{1}(j)}{D_{2}(j)}$ with $D_1$ represents the true group distribution between $G_0$ and $G_1$ in top-K recommendation list, and $D_2=[\frac{|G_0|}{|\mathcal{I}|},\frac{|G_1|}{|\mathcal{I}|}]$ represents their ideal distribution of the overall population.


\subsection{Experimental Results}
The major experimental results are shown in Table \ref{tab:result}, besides, we also plot the approximate Pareto frontier between NDCG and Long-tail Rate (namely, 1-Popularity Rate) in Fig. \ref{fig:pareto_frontier}. 
We analyze and discuss the results in terms of the following perspectives.

\subsubsection{Recommendation Performance}
For recommendation perfor-mance, we compare MoFIR-1.0 with MF, BPR, NGCF, and LIRD based on $Recall@k$, $F1@k$ and $NDCG@k$ and provide the these results of the recommendation performance in Table \ref{tab:result}.
Among all the baseline models, we can see that all sequential recommendation methods (LIRD, MoFIR-1.0) are much better than the traditional method, which demonstrates the superiority of sequential recommendation on top-K ranking tasks.
Specifically, LIRD is the strongest baseline in all three datasets on all performance metrics: when averaging across recommendation lengths LIRD achieves 41.28\% improvement than MF, 27.08\% improvement than BPR-MF, and 8.97\% improvement than NGCF. 

Our MoFIR approach achieves the best top-K recommendation performance against all baselines on all datasets: when averaging across three recommendation lengths on all performance metrics, MoFIR gets 41.40\% improvement than the best baseline on Movielens100K; MoFIR gets 46.45\% improvement than LIRD on Ciao; and MoFIR gets 18.98\% improvement than LIRD on Etsy.
These above observations imply that the proposed method does have the ability to capture the dynamic nature in user-item interactions, which results in better recommendation results. 
Besides, unlike LIRD, which only concatenates user and item embeddings together, MoFIR uses several GRU layers to better capture the sequential information in user history, which benefits the model performance.

\subsubsection{Fairness Performance}
For fairness performance, we compare MoFIRs with FOE-based methods and MFR-based methods based on $KL\ Divergence@k$ and $Popularity\ Rate@k$, which are also shown in Table \ref{tab:result}.
It is easy to find that there does exist a trade-off between the recommendation performance and the fairness performance, which is understandable, as most of the long-tail items have relatively fewer interactions with users.
When comparing the baselines, we can easily find that MFR is able to achieve better trade-off than FOE as it is also a multi-objective optimization method.

From Table \ref{tab:result}, MoFIR is able to adjust the degree of trade-off between utility and fairness through simply modifying the weight of recommendation utility objective.
It is worth noting that MoFIR-0.1 can always closely achieve the ideal distribution as its $KL$s are close to zero.
In Table \ref{tab:result}, we can find that even MoFIR has the similar performance of fairness with other baselines, it can still achieve much better recommendation performance (for example, BPR-FOE and MoFIR-0.5 in Movielens100k or NGCF-FOE and MoFIR-0.5 in Ciao or MF-MFR and MoFIR-0.5 in Etsy), which indicates its capability of finding better trade-off.

\subsubsection{Fairness-Utility Trade-off}
We only compare MoFIR with MFR, since FOE is a post-processing method, which doesn't optimize the fairness-utility trade-off.
In order to better illustrate the trade-off between utility and fairness, we fix the length of the recommendation list at 20 and plot $NDCG@20$ against $Longtail\ Rate$ in Fig. \ref{fig:pareto_frontier} for all datasets, where $Longtail\ Rate$ equals to one minus $Popularity\ Rate$.
Each blue point is generated by simply changing the input weights to the fine-tuned MoFIR, while each orange point is generated by running the entire MFR optimization.
The clear margin distance between the blue points' curve (Approximate Pareto frontier) and the orange points' curve demonstrates the great effectiveness of MORL compared with traditional multi-objective optimization method in recommendation.

\vspace{-1pt}
\section{Conclusion}
In this work, we achieve the approximate Pareto efficient trade-off between fairness and utility in recommendation systems and characterize their  Pareto Frontier in the objective space in order to find solutions with different level of trade-off.
We accomplish the task by proposing a fairness-aware recommendation framework using multi-objective reinforcement learning (MORL) with linear preferences, called MoFIR, which aims to learn a single parametric representation for optimal recommendation policies over the space of all possible preferences.
Experiments across three different datasets demonstrate the effectiveness of our approach in both fairness measures and recommendation performance.

\section*{Acknowledgments}
We gratefully acknowledge the valuable cooperation of Runzhe Yang from Princeton University and Shuchang Liu from Rutgers University. 


\balance
\bibliographystyle{ACM-Reference-Format}
\bibliography{sample-base}


\end{document}